\documentclass[conference]{IEEEtran}
\IEEEoverridecommandlockouts

\usepackage{cite}
\usepackage{amsmath,amssymb,amsfonts}
\usepackage{algorithmic}
\usepackage{graphicx}
\usepackage{textcomp}
\usepackage{xcolor}
\usepackage{url}

\usepackage{bm}
\usepackage{authblk}

\newtheorem{lemma}{\underline{Lemma}}
\newtheorem{proposition}{\underline{Proposition}}
\newtheorem{remark}{\underline{Remark}}
\def\BibTeX{{\rm B\kern-.05em{\sc i\kern-.025em b}\kern-.08em
	T\kern-.1667em\lower.7ex\hbox{E}\kern-.125emX}}

\begin{document}

\title{Multi-User ISAC with Heterogeneous Unknown Parameters: Optimal Beamforming based on Distribution Information\\
}
\author[*]{Chan Xu}
\author[$\dagger$]{Shuowen Zhang}
\affil[*]{South-Central Minzu University, Wuhan, China, xuchan@mail.scuec.edu.cn}
\affil[$\dagger$]{The Hong Kong Polytechnic University, Hong Kong SAR, China, shuowen.zhang@polyu.edu.hk}
\maketitle

\begin{abstract}
	This paper studies an integrated sensing and communication (ISAC) system where a multi-antenna base station (BS) communicates with multiple single-antenna users in the downlink and senses the \textit{unknown} and \textit{random} angle information of a target based on its prior distribution information and the received echo signals. We focus on a challenging scenario with \emph{heterogeneous unknown parameters} where the target's reflection coefficient is also \emph{unknown} with no prior information. We consider a general transmit beamforming structure with both communication beams and \textit{dedicated sensing} beams, where the communication users can cancel the interference caused by the pre-determined sensing signals. By adopting the \textit{periodic posterior Cram\'{e}r-Rao bound (PCRB)} to quantify a lower bound of the mean-cyclic error (MCE) for sensing the periodic angle parameter, we optimize the transmit beamforming to minimize the periodic PCRB, subject to individual communication user rate constraints, which is a non-convex problem. By leveraging the semi-definite relaxation (SDR) technique and Lagrange duality theory, we derive the \textit{optimal solution} and prove that \textit{at most one} dedicated sensing beam is needed. Numerical results validate our analysis and effectiveness of the proposed beamforming design.
\end{abstract}

 \vspace{-2mm}
\section{Introduction}
 \vspace{-1mm}
Sixth-generation (6G) wireless networks will incorporate sensing as a new function, which enables various new applications such as autonomous driving, low-altitude UAV management, and industrial digital twins \cite{6G}. Integrated sensing and communication (ISAC) will support both functionalities via a unified infrastructure, thereby enhancing spectral efficiency and reducing deployment costs \cite{ISAC_survey1}. With ISAC, base stations (BSs) can perform downlink communication and estimate parameters of sensing targets via received echo signals, for which transmit beamforming optimization is very critical.

Among the existing works on transmit beamforming optimization, Cram\'{e}r-Rao bound (CRB) has been commonly adopted as the sensing performance metric \cite{jointDesign_CRB1,jointDesign_CRB2,jointDesign_CRB3}, which is a lower bound of the sensing mean-squared error (MSE) \cite{CRB}. However, CRB is a function of the \emph{exact values} of the parameters to be sensed, which are generally \emph{unknown} prior to beamforming and sensing. On the other hand, the probability density function (PDF) of such parameters can be known \emph{a priori} based on historic data or statistical analysis \cite{ISIT,YaoJiayi,JSAC_MIMO, WangYizhuo,liu2024ris,zheng2025beyond,Nadim,HouKaiyue2,ISIT24,YaoJiayi2,Yu_Bound2,Yuan_PCRB,Yu,Caire,HouKaiyue3}. Based on the available prior information, \emph{posterior Cram\'{e}r-Rao bound (PCRB)} can be derived to quantify a lower bound of the MSE \cite{VanTrees}, which is only dependent on the PDF and tight in moderate-to-high signal-to-noise ratio (SNR) regimes.

\begin{figure}[t]
	\centering
	\includegraphics[width=3.1in]{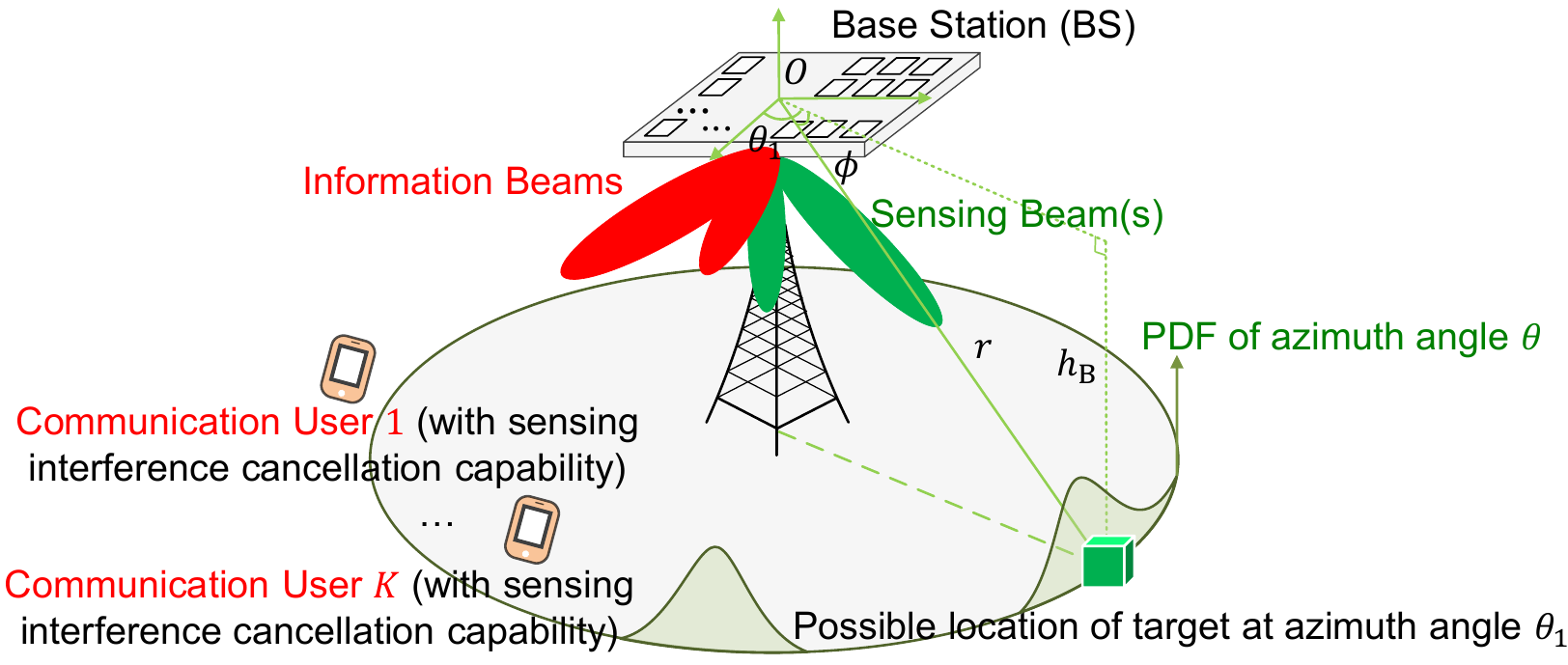}
	\vspace{-3mm}
	\caption{Illustration of multi-user ISAC with sensing interference cancellation capability and dedicated sensing beams.}\label{fig_system}
\vspace{-1mm}
\end{figure}

Prior studies on PCRB-based beamforming mainly considered dual-function beams for both sensing and communication \cite{JSAC_MIMO,WangYizhuo,liu2024ris,zheng2025beyond,Nadim}. Since sensing and communication have drastically different objectives, and the locations of communication users and sensing target are generally different, sending only dual-functional beams may be strictly sub-optimal in terms of the sensing-communication trade-off. Moreover, in a multi-user system, the existence of multi-user interference further limits the design flexibility of the communication beams. The beamforming optimization in the general case with \emph{dedicated sensing beams} was recently studied in \cite{ISIT24,Yu_Bound2,YaoJiayi2}. Specifically, when all the unknown parameters in the system are all \emph{random} with \emph{known distribution}, i.e., being \emph{homogeneous}, \cite{YaoJiayi2,Yu_Bound2} derived various bounds on the number of beams needed. However, for the case with \emph{heterogeneous} unknown parameters where some parameters are \emph{random} with \emph{known distribution}, while others are \emph{deterministic} or with no known prior information, the PCRB characterization will be more complicated, for which only \cite{ISIT24} studied a single-user case. Note that this is a highly practical case and corresponds to various scenarios where there is no information about the target's reflection coefficient, non line-of-sight (LoS) clutter, etc., while the corresponding parameters still need to be jointly estimated with the desired parameters (e.g., angle/range).

This paper aims to make the first attempt to study the transmit beamforming optimization in a multi-user ISAC system with \emph{heterogeneous unknown parameters}, under a general structure with possible dedicated sensing beams, as illustrated in Fig. \ref{fig_system}. The BS sends downlink signals to perform multi-user communication and sense the \emph{unknown} and \emph{random} angle of a point target based on its reflected echo signals and perfect prior PDF of the angle; while the target reflection coefficient is also unknown and coupled in the echo signals, for which no information is available. Specifically, since the sensing signals can be generated by pseudo random coding \cite{RadarSignal} as pre-determined sequences, we assume the communication users can cancel the interference from sensing signals perfectly. Due to the periodicity of the angle, we adopt periodic PCRB as the sensing performance metric and formulate the beamforming optimization problem to minimize the periodic PCRB subject to the individual rate requirement at each communication user. Despite the non-convexity of the problem, we leverage the semi-definite relaxation (SDR) technique to obtain the optimal solution by proving its tightness. By leveraging Lagrange duality theory, we further prove that \textit{at most one} dedicated sensing beam is needed, which is validated via numerical results.
 \vspace{-2mm}
\section{System Model }\label{Sec_sys}
Consider a multi-user ISAC system which consists of a BS equipped with $N_\mathrm{T}\geq 1$ transmit antennas and $N_\mathrm{R}\geq 1$ co-located receive antennas, $K\leq N_\mathrm{T}$ single-antenna communication users, and a point ground target with \emph{unknown} and \emph{random} location to be sensed, as illustrated in Fig. \ref{fig_system}. Specifically, the antennas at the BS are deployed under a uniform planar array (UPA) configuration. We assume that the distance (range) $r$ and elevation angle $\phi$ of target with respect to the BS are known, and the azimuth angle of target denoted by $\theta\in[-\pi,\pi)$ is the \emph{unknown} and \emph{random} location parameter to be sensed. The PDF of $\theta$ denoted by $p_\Theta(\theta)$ is assumed to be known \emph{a priori} from historic data or target appearance pattern \cite{ISIT,YaoJiayi,JSAC_MIMO, WangYizhuo,liu2024ris,zheng2025beyond,Nadim,HouKaiyue2,ISIT24,YaoJiayi2,Yu_Bound2,Yuan_PCRB,Yu,Caire,HouKaiyue3}. The BS sends downlink signals to deliver information to multiple users, and estimate the unknown parameter $\theta$ based on the target-reflected echo signals and the prior distribution information $p_\Theta(\theta)$.

We consider an LoS channel between the BS and the target. Let $\bm{a}(\phi,\theta)\in\mathbb{C}^{N_\mathrm{T}\times1}$ and $\bm{b}(\phi,\theta)\in\mathbb{C}^{N_\mathrm{R}\times1}$ denote the steering vectors of the transmit and receive antenna arrays, respectively. Let $\alpha =\alpha_\mathrm{R}+j\alpha_\mathrm{I}\in\mathbb{C}$ denote the overall complex reflection coefficient of the target, which consists of the round-trip path loss and radar cross-section (RCS) coefficient. The overall channel from the BS transmitter to the BS receiver via target reflection is thus given by $\alpha\bm{b}(\phi,\theta)\bm{a}^H(\phi,\theta)$.

Due to the unknown RCS of the target, the reflection coefficient $\alpha$ is also \emph{unknown}. Particularly, we focus on a challenging case where \emph{no prior information} on $\alpha$ is available (e.g., $\alpha$ may be a deterministic parameter or a random parameter with no known information). On the other hand, different from the desired parameter $\theta$, $\alpha$ is a \emph{nuisance parameter} \cite{shen}. This thus corresponds to a complex yet practical sensing scenario with \emph{heterogeneous parameters} with known/unknown prior information and desired/nuisance properties.

We consider linear transmit beamforming design under a general structure where the transmit signal is the superposition of $K$ \emph{communication beams} and one or multiple \emph{dedicated sensing beams}. We focus on $L\geq 1$ symbol intervals, during which both the target's location and the communication channel between the BS and the users remain unchanged. Let $\bm{w}_k\in\mathbb{C}^{N_\mathrm{T}\times 1}$ denote the communication beamforming vector for the $k$-th single-antenna user. Let $\bm{S}=[\bm{s}_1,...,\bm{s}_{N_\mathrm{T}}]\in\mathbb{C}^{N_\mathrm{T}\times N_\mathrm{T}}$ denote the sensing beamforming matrix to be optimized, with $\bm{s}_i\in\mathbb{C}^{N_\mathrm{T}\times1}$ denoting the beamforming vector for the $i$-th sensing symbol. The minimum number of non-zero columns in the optimized $\bm{S}$ represents the number of dedicated sensing beams needed. In each $l$-th symbol interval, let $\bm{c}_l=[c_{l,1},...,c_{l,K}]^T\sim \mathcal{CN}(\bm{0},\bm{I}_K)$ denote the information symbol vector for $K$ communication users and $\bm{\upsilon}_l\in\mathbb{C}^{N_\mathrm{T}\times 1}\sim \mathcal{CN}(\bm{0},\bm{I}_{N_\mathrm{T}})$ denote the dedicated sensing signals generated by pseudo random coding, where the elements in $\bm{c}_l$ and $\bm{v}_l$ are also independent of each other. The transmitted signal vector in the $l$-th symbol interval is thus given by
$\bm{x}_l=\sum_{k=1}^{K}\bm{w}_kc_{l,k}+\bm{S}\bm{\upsilon}_l,\quad \forall l$.
The transmit covariance matrix is given by $\bm{R}_X=\mathbb{E}[\bm{x}_l\bm{x}_l^H]=\sum_{k=1}^{K}\bm{w}_k\bm{w}_k^H+\bm{S}\bm{S}^H$. Let $P$ denote the transmit power budget at the BS transmit antennas, which yields $\sum_{k=1}^{K}\|\bm{w}_k\|^2+\mathrm{tr}(\bm{SS}^H)\leq P$.

\vspace*{-2mm}
\subsection{Sensing with Heterogeneous Unknown Parameters}
\vspace{-1mm}
Based on the above model, the received echo signal at the BS receive antennas in the $l$-th symbol interval is given by
\begin{align}
	\vspace{-4.5mm}
	\bm{y}_l^\mathrm{S}=\alpha\bm{b}(\phi,\theta)\bm{a}^H(\phi,\theta)\bm{x}_l+{\bm{n}_l^\mathrm{S}},\quad \forall l,
	\vspace{-4mm}
\end{align}
where $\bm{n}_l^\mathrm{S}\sim\mathcal{CN}(\bm{0},\sigma_\mathrm{S}^2\bm{I}_{N_\mathrm{R}})$  denotes the circularly symmetric complex Gaussian (CSCG) noise vector over the BS receive antennas in the $l$-th symbol interval, with $\sigma_\mathrm{S}^2$ denoting the average noise power. Let $\bm{\zeta}=[\theta,\alpha_{\mathrm{R}},\alpha_{\mathrm{I}}]^T$ denote the collection of all the unknown parameters, where $\theta$ is the desired parameter with prior distribution information, while $\alpha_{\mathrm{R}}$ and $\alpha_{\mathrm{I}}$ are nuisance parameters without any information. Nevertheless, all the parameters in $\bm{\zeta}$ need to be jointly estimated to obtain an accurate estimate of the desired parameter $\theta$, based on the observations in $L$ symbol intervals, $\bm{Y}=[{\bm{y}_1^\mathrm{S}},...,{\bm{y}_L^\mathrm{S}}]$, and the prior PDF information of $\theta$, $p_\Theta(\theta)$.

Motivated by the \emph{periodic} feature of the angle parameter $\theta$, we adopt the MCE to quantify the sensing performance of $\theta$, which takes the $2\pi$-periodicity of $\theta$ into account and is more accurate compared to the MSE \cite{ISIT24}. Specifically, the MCE is given by $\mathrm{MCE}=2-2\mathbb{E}_{\bm{Y}, \bm{\zeta}}[\cos(\hat{\theta}-\theta)]$, where $\hat{\theta}$ denotes the estimate of $\theta$. However, MCE is difficult to be characterized explicitly, thus we adopt the periodic PCRB as a lower bound of the MCE as the sensing performance metric for our beamforming optimization, which is tight in moderate-to-high SNR regimes \cite{VanTrees,periodic_PCRB}.  Since all the parameters in $\bm{\zeta}$ need to be jointly estimated, we let $\bm{F}\in \mathbb{C}^{3\times 3}$ denote the periodic posterior Fisher information matrix (PFIM) for $\bm{\zeta}$, based on which the periodic PCRB of the MCE for the desired sensing parameter $\theta$ can be shown to be given by $\mathrm{PCRB}^{\mathrm{P}}_{\theta}\!=\!2-2\left(1\!+\![\bm{F}^{-1}]_{1,1}\right)^{-\frac{1}{2}}$ \cite{periodic_PCRB}. Note that the periodic PFIM for this case with heterogeneous parameters can be derived following the rationale in \cite{shen}. Based on this, the periodic PCRB can be further derived as \cite{ISIT24}
\begin{align}\label{PCRB}
	\vspace*{-6mm}
	&\mathrm{PCRB}^{\mathrm{P}}_{\theta}=2-\frac{2}{(1+ [\bm{F}^{-1}]_{1,1})^{\frac{1}{2}}}=2-\nonumber\\[-1mm]
    &\frac{2}{\sqrt{1\!+\!\frac{1}{\mathbb{E}_\theta\left[\!\left(\frac{\partial\ln(\hat{p}_\Theta(\theta))}{\partial\theta}\right)^2\!\right]+\frac{2|\alpha|^2L}{\sigma_{\mathrm{S}}^2}\big(\!\mathrm{tr}\left(\bm{A}_1\!\bm{R}_X\right)-\frac{\left|\mathrm{tr}(\!\bm{A}_2\!\bm{R}_X\!)\right|^2}{\mathrm{tr}(\!\bm{A}_3\!\bm{R}_X\!)}\!\big)}}},	
\end{align}
where $\hat{p}_\Theta(\theta)\!\!\!=\!\!\!{p}_\Theta(\theta\!\!-\!\!2\pi\lfloor\frac{\theta+\pi}{2\pi}\rfloor),\theta\!\!\!\in\!\!\! \mathbb{R}$, $\bm{A}_1\!\!\!=\!\!\!\int_{-\pi}^\pi(\|\dot{\bm{b}}(\phi,\!\theta)\|^2\bm{a}(\phi,\!\theta)\bm{a}^H(\phi,\theta)\!+\!N_\mathrm{R}\dot{\bm{a}}(\phi,\!\theta)\dot{\bm{a}}^H\!(\phi,\!\theta))p_\Theta(\theta)d\theta$, $\bm{A}_2\!=\!N_\mathrm{R}\int_{-\pi}^\pi\dot{\bm{a}}(\phi,\theta)\bm{a}^H(\phi,\theta)p_\Theta(\theta)d\theta$, and $\bm{A}_3\!=\!N_\mathrm{R}\int_{-\pi}^\pi\bm{a}(\phi,\theta)\bm{a}^H(\phi,\theta)p_\Theta(\theta)d\theta$, with $\dot{\bm{a}}(\phi,\theta)$, $\dot{\bm{b}}(\phi,\theta)$ denoting the derivatives of ${\bm{a}}(\phi,\theta)$, ${\bm{b}}(\phi,\theta)$ with respect to $\theta$. Detailed proof is omitted due to limited space, which can be found in Appendix A of our prior work \cite{ISIT24}.
\begin{remark}\emph{(Difference between PCRBs with heterogeneous vs. homogeneous unknown parameters)}.
Notice from (\ref{PCRB}) that $\mathrm{PCRB}^{\mathrm{P}}_{\theta}$ is in a complex form with layers of fractions with respect to $\bm{R}_X$. This is fundamentally different from the homogeneous case with prior distribution information for all unknown parameters \cite{YaoJiayi2,Yu_Bound2}. For example, $\mathrm{tr}(\bm{A}_1\!\bm{R}_X)-\frac{\left|\mathrm{tr}(\!\bm{A}_2\!\bm{R}_X\!)\right|^2}{\mathrm{tr}(\!\bm{A}_3\!\bm{R}_X\!)}$ is replaced by the trace of a linear function of $\bm{R}_X$ since the effect of $\alpha$ can be ``averaged out'' by leveraging its distribution information and a mild condition in \cite{YaoJiayi2}. Moreover, the periodic PCRB for the single desired sensing parameter in the heterogeneous case is determined by \emph{only one} element in the inverse PFIM $\bm{F}^{-1}$, instead of \emph{all} diagonal elements in $\bm{F}^{-1}$ \cite{Yu_Bound2} where all parameters are treated as desired parameters.
\end{remark}
\vspace*{-2mm}
\subsection{Communication with Sensing Interference Cancellation}
\vspace*{-1mm}
Let $\bm{h}_k^H\in\mathbb{C}^{1\times N_\mathrm{T}}$ denote the BS-user channel of the $k$-th user, which is assumed to be known perfectly at both the BS and the user. The received signal at the $k$-th communication user receiver in each $l$-th symbol interval is given by
\begin{align}\label{y_C}
	\vspace{-2mm}		\!y^\mathrm{C}_{l,k}\!=\!\bm{h}_k^H\bm{w}_kc_{l,k}\!+\!\!\sum_{i{\neq}k}^{K}\!\bm{h}_k^H\bm{w}_ic_{l,i}\!+\!\bm{h}_k^H\bm{S}\bm{\upsilon}_l\!+\!n^\mathrm{C}_{l,k},\quad \forall l,
	\vspace{-2mm}
\end{align}
where $n^\mathrm{C}_{l,k}\sim\mathcal{CN}(0,\sigma_k^2)$ denotes the CSCG noise at the $k$-th user's receive antenna in the $l$-th symbol interval, with $\sigma_k^2$ denoting the average noise power. Note that extra signals from dedicated sensing beams, $\bm{h}_k^H\bm{S}\bm{\upsilon}_l$'s, are received at each $k$-th user, which leads to undesired interference. However, since the dedicated sensing signals are pre-determined sequences generated by pseudo random coding, they can be known \emph{a priori} at both the BS and each user, together with the user's own channel and the designed sensing beamforming matrix. With such knowledge, we assume the communication users possess the capability of cancelling the interference caused by sensing signals perfectly. The achievable rate of each $k$-th user is thus given by
\begin{align}\label{capacity2}
	\vspace{-2mm}
	R_k=\log_2\left(1+\frac{|\bm{h}_k^H\bm{w}_k|^2}{\sum_{i=1,i\neq{k}}^{K}|\bm{h}_k^H\bm{w}_i|^2+\sigma_k^2}\right)
	\vspace{-1mm} 	
\end{align}
in bits per second per Hertz (bps/Hz).

Note that both the sensing and communication performance metrics are critically determined by the transmit beamforming, whose optimization will be studied in the following to achieve an optimal sensing-communication trade-off. We will also reveal \emph{if} and \emph{how many} dedicated sensing beams are needed in the optimal solution in this challenging scenario with multi-user communication and heterogeneous unknown parameters.
\vspace*{-6mm}
\section{Problem Formulation}\label{Sec_pro}
\vspace*{-1mm}
We aim to optimize the transmit beamforming with potential dedicated sensing beams to minimize the sensing periodic PCRB, subject to the individual communication rate constraint at each $k$-th communication user specified by rate target $\bar{R}_k$ bps/Hz. The optimization problem is formulated as
\begin{align} \label{P1}
	\!\!\!	\vspace{-2mm}
	\mbox{(P1)}\mathop{\mathrm{min}}_{\{\bm{w}_k\}_{k=1}^{K},\bm{S}}\  &\mathrm{PCRB}_{\theta}^\mathrm{P}\\[-2mm]
	\mathrm{s.t.}\
	&\log_2\Bigg(1\!+\!\frac{|\bm{h}_k^H\bm{w}_k|^2}{\sum\limits_{i\neq{k}}|\bm{h}_k^H\bm{w}_i|^2+\sigma_k^2}\Bigg)\geq\bar{R}_k,\;\forall k\label{P1_C2}\\[-2mm]
	&\sum_{k=1}^{K}\|\bm{w}_k\|^2+\mathrm{tr}(\bm{SS}^H)\leq P. \label{P1_C3}
	\vspace{-2mm}
\end{align}
Note that Problem (P1) is a non-convex optimization problem, due to the non-convex and complex fractional structure of the objective function as well as the non-convex constraints in (\ref{P1_C2}). Particularly, this problem is highly non-trivial and more complicated than that for single-user ISAC \cite{ISIT24}, since the existence of multi-user interference under the rate constraints in (\ref{P1_C2}) significantly reduces the feasible space of the communication beamformers and potentially necessitates more sensing beams; while more sensing beams are indeed allowed since the sensing interference can be cancelled at the communication users. Moreover, due to the heterogeneous unknown parameters involved in this challenging system, existing results with homogeneous parameters \cite{YaoJiayi2,Yu_Bound2} are not applicable. In the following, we address these new challenges by deriving the optimal solution to (P1) and reveal the number of dedicated sensing beams needed.
\vspace*{-2mm}
\section{Optimal Solution to Problem (P1)}\label{Sec_sol}
\vspace*{-1mm}
\subsection{ Equivalent Problem Transformation}
\vspace{-1mm}
Firstly, we note that the minimization of $\mathrm{PCRB}_{\theta}^\mathrm{P}$ is equivalent to the maximization of $\mathrm{tr}(\bm{A}_1\bm{R}_X)-\frac{\left|\mathrm{tr}(\bm{A}_2\bm{R}_X)\right|^{2}}{\mathrm{tr}(\bm{A}_3\bm{R}_X)}$. By introducing auxiliary variables $t$ and $\gamma_k=2^{\bar{R}_k}-1,\ \forall k$, (P1) is equivalently transformed to the following problem:
\begin{align}
	\vspace{-2mm}
	\!\!\!\mbox{(P2)}\mathop{\mathrm{max}}_{t,\{\bm{w}_k\}_{k=1}^{K},\bm{S}}\ &t\\[-1.5mm]
	\mathrm{s.t.}\ & \mathrm{tr}(\bm{A}_1\bm{R}_X)-\frac{\left|\mathrm{tr}(\bm{A}_2\bm{R}_X)\right|^{2}}{\mathrm{tr}(\bm{A}_3\bm{R}_X)}\geq t \label{constraint_t}\\[-1.5mm] &|\bm{h}_k^H\bm{w}_k|^2\!-\!\gamma_k\sum_{i\neq{k}}|\bm{h}_k^H\bm{w}_i|^2\geq\gamma_k\sigma_k^2,\;\forall{k}\!\\[-1.5mm]
	&\sum_{k=1}^{K}\|\bm{w}_k\|^2+\mathrm{tr}(\bm{SS}^H)\leq P.
	\vspace{-2mm}
\end{align}

Then, we adopt SDR to solve (P2) by defining $ \bm{W}_k\triangleq\bm{w}_k\bm{w}_k^H $ with $\mathrm{rank}(\bm{W}_k)=1,\forall k$, and $\bm{W}_\mathrm{S}\triangleq \bm{SS}^H $, which yields  $\bm{R}_X=\sum\limits_{k=1}^{K}\bm{W}_k+\bm{W}_\mathrm{S}$. Based on the Schur complement theory \cite{schur}, (\ref{constraint_t}) is equivalent to $\bm{B}\left(t,\{\bm{W}_k\}_{k=1}^{K},\bm{W}_\mathrm{S}\right)\overset{\Delta}{=}\left[\begin{array}{ll}
	\mathrm{tr}(\bm{A}_1\bm{R}_X)-t&\mathrm{tr}(\bm{A}_2\bm{R}_X)\\
	\mathrm{tr}(\bm{A}_2^H\bm{R}_X)&\mathrm{tr}(\bm{A}_3\bm{R}_X)
\end{array}\right]\succeq \bm{0}$. Thus, (P2) and (P1) are equivalent to the following problem with additional constraints $\mathrm{rank}(\bm{W}_k)=1,\forall k$:
\begin{align} \label{P2-R}
	\vspace{-2mm}
	\!\!\!\!\mbox{(P2-R)}\mathop{\mathrm{max}}_{ t,\{\!\bm{W}_k\!\}_{k=1}^{K},\bm{W}_\mathrm{S}}\ &t\\[-1mm]
	\mathrm{s.t.}\quad &\bm{B}(t,\{\bm{W}_k\}_{k=1}^{K},\bm{W}_\mathrm{S})\succeq\bm{0}\label{P2R_C1}\\[-1mm]
	&\bm{h}_k^H(\bm{W}_k\!-\!\gamma_k\sum_{i\neq k}\bm{W}_i)\bm{h}_k\!\geq\!\gamma_k\sigma_k^2,\;\forall k \!\!\!\label{P2R_C3}\\[-2mm]
	&\mathrm{tr}\Big(\sum_{k=1}^{K}\bm{W}_k+\bm{W}_\mathrm{S}\Big)\leq P \label{P2R_C4}
	\vspace{-2mm}
\end{align}
\begin{align}
    &\!\!\!\bm{W}_k\succeq\bm{0},\quad\forall k  \label{P2R_C5}\quad\quad\\[-1.5mm]
	&\!\!\!\bm{W}_\mathrm{S}\succeq\bm{0}. \label{P2R_C6}
\end{align}
(P2-R) is a convex semi-definite program (SDP), for which the optimal solution can be obtained via the interior-point method \cite{CVX} or existing software, e.g., CVX \cite{cvxtool}. Note that if the optimal solution to (P2-R) satisfies the rank-one constraints, the SDR relaxation from (P2) to (P2-R) is \emph{tight}, and an optimal solution to (P2) and (P1) can be obtained. In the following, we analyze the tightness of the SDR, and leverage Lagrange duality theory to unveil tight bounds on the number of dedicated sensing beams needed in the optimal solution.
\vspace*{-8mm}
\subsection{Optimal Solution to (P1)}
\vspace{-3mm}
We first prove the tightness of SDR from (P1) and (P2) to (P2-R) via the following proposition.
\begin{proposition}\label{prop_Communication_Beam}
	Given any optimal solution to (P2-R) denoted by $(t^\star,\{\bm{W}^\star_k\}_{k=1}^{K},\bm{W}^\star_\mathrm{S})$, $(t^\star,\{\bar{\bm{W}}^\star_k\}_{k=1}^{K},\bar{\bm{W}}^\star_\mathrm{S})$ is also an optimal solution to (P2-R), with
	\begin{align}
\vspace{-2mm}		\bar{\bm{W}}^\star_k&=\bm{W}^\star_k\bm{h}_k\bm{h}_k^H\bm{W}^\star_k/(\bm{h}_k^H\bm{W}^\star_k\bm{h}_k),\quad\forall k,\label{construct_Wu}\\[-1mm]	\bar{\bm{W}}^\star_\mathrm{S}&=\bm{W}^\star_\mathrm{S}+\sum_{k=1}^{K}(\bm{W}^\star_k-\bar{\bm{W}}_k^\star).\label{construct_WS}
\vspace{-2mm}	
\end{align}
\end{proposition}
\begin{IEEEproof}[Proof Sketch]
    $(t^\star,\{\bar{\bm{W}}^\star_k\}_{k=1}^{K},\bar{\bm{W}}^\star_\mathrm{S})$ can be shown to be feasible for (P2-R) and achieve the optimal value of (P2-R). Moreover, the power constraint in (\ref{P2R_C4}) holds with equality at the optimal solution, as otherwise allocating more power to sensing beams always improves the periodic PCRB. Please refer to Appendix A for detailed proof.
\end{IEEEproof}

Proposition \ref{prop_Communication_Beam} proves that we can always find an optimal solution to (P2-R) satisfying the rank-one constraints on $\bm{W}_k$'s. Hence the SDR from (P1) and (P2) to (P2-R) is \emph{tight}. The optimal solution to (P1) can be found by first solving (P2-R) and then decomposing $\bar{\bm{W}}^\star_k$ and $\bar{\bm{W}}^\star_\mathrm{S}$.
\vspace*{-8mm}
\subsection{Bound on the Number of Dedicated Sensing Beams Needed}
\vspace{-3mm}
Note that although inclusing more sensing beams enables more design flexibility, it also requires higher cost and complexity at the transmitter. In this subsection, we aim to unveil the minimum number of dedicated sensing beams that is sufficient for achieving the optimal solution to (P1), namely, the minimum rank of $\bm{W}_{\mathrm{S}}$ among all the optimal solutions.

We first note that if (P2-R) is a separable SDP, the rank of the optimal solution to each positive semi-definite matrix variable can be bounded by the number of constraints, and the rank-reduction method in \cite{Palomar} can be readily applied, which is the case with homogeneous unknown parameters in \cite{YaoJiayi2}. However, for our considered case with heterogeneous unknown parameters, due to the non-linearly coupled variables $t$, $\bm{W}_k$'s, and $\bm{W}_{\mathrm{S}}$ in (\ref{P2R_C1}), (P2-R) is not a separable SDP, for which the general bound and rank-reduction method in \cite{Palomar} cannot be applied. In the following, we leverage Lagrange duality theory \cite{CVX} to design a \emph{tailored} rank-reduction method.

Denote $\bm{Z}_B=[z_1,z_2;z_2^*,z_3]\succeq\bm{0}$, $\bm{\mu}=[\mu_1,...,\mu_{K}]^T\succeq\bm{0}$, $\eta\geq0$, $\bm{Z}_k\succeq\bm{0},\ \forall k$, and $\bm{Z}_\mathrm{S}\succeq \bm{0}$ as the dual variables associated with the constraints in (\ref{P2R_C1})-(\ref{P2R_C6}), respectively. The Lagrangian of Problem (P2-R) is given by
\begin{align}\label{P4_LA}
\vspace{-2mm} &\mathcal{L}\left(t,\{\bm{W}_k\}_{k=1}^{K},\bm{W}_\mathrm{S},\bm{Z}_B,\bm{\mu},\eta,\{\bm{Z}_k\}_{k=1}^{K},\bm{Z}_\mathrm{S}\right)\nonumber\\[-2mm]	=&t\!+\!\mathrm{tr}\left(\bm{Z}_B\bm{B}\left(t,\{\bm{W}_k\}_{k=1}^{K},\bm{W}_\mathrm{S}\right)\!\right)\!\!-\!\eta\Big(\!\mathrm{tr}\Big(\!\sum_{k=1}^{K}\bm{W}_k\!\!+\!\!\bm{W}_\mathrm{S}\Big)\!\!-\!\!P\Big)\nonumber\\[-2mm]
	&+\sum_{k=1}^{K}\mu_k(\bm{h}_k^H(\bm{W}_k-\gamma_k\sum_{i\neq k}\bm{W}_i)\bm{h}_k-\gamma_k\sigma_k^2)\nonumber\\[-2mm]
	&+\sum_{k=1}^{K}\mathrm{tr}(\bm{Z}_k\bm{W}_k)\!+\!\mathrm{tr}(\bm{Z}_\mathrm{S}\!\bm{W}_\mathrm{S}),
\vspace{-3mm}
\end{align}
where $ \mathrm{tr}\left(\bm{Z}_B \bm{B}\left(t,\{\bm{W}_k\}_{k=1}^{K},\bm{W}_\mathrm{S}\right)\right)=\mathrm{tr}(\bm{D}(\sum_{k=1}^{K}\bm{W}_k+ \bm{W}_\mathrm{S}))-z_1t$ with $\bm{D}=z_1\bm{A}_1+z_2\bm{A}^H_2+z_2^*\bm{A}_2+z_3\bm{A}_3$.

Define the optimal primal and dual solutions for (P2-R) as $t^\star$, $\{\bm{W}_k^\star\}_{k=1}^K$, $\bm{W}_\mathrm{S}^\star$, and $\bm{Z}_B^\star=[z_1^\star,z_2^\star;z_2^{\star^*},z_3^\star]$, $\bm{\mu}^\star=[\mu_1^\star,...,\mu^\star_{K}]^T$, $\eta^\star$, $\{\bm{Z}_k^\star\}_{k=1}^K$, $\bm{Z}_\mathrm{S}^\star$, respectively. For simplicity, we use $\mathcal{L}^\star$ to denote $\mathcal{L}(t^\star,\{\bm{W}^\star_k\}_{k=1}^{K},\bm{W}^\star_\mathrm{S},\bm{Z}^\star_B,\bm{\mu}^\star,\eta^\star,\{\bm{Z}^\star_k\}_{k=1}^{K},\bm{Z}^\star_\mathrm{S})$. Note that the strong duality holds for (P2-R). The KKT conditions \cite{CVX} of (P2-R) are given by
\begin{align}
\vspace{-2mm}	\mathrm{tr}\left(\bm{Z}_B^\star\bm{B}\left(t^\star,\{\bm{W}_k^\star\}_{k=1}^{K},\bm{W}^\star_\mathrm{S}\right)\right)=&0 \label{KKT1}\\[-1.5mm]	\mu^\star_k(\bm{h}_k^H(\bm{W}^\star_k\!-\!\gamma_k\sum_{i\neq{k}}\bm{W}^\star_i)\bm{h}_k-\gamma_k\sigma_k^2)=&0,\quad\forall k\label{KKT_rate2}\\[-1.5mm]
	\eta^\star\Big(\mathrm{tr}\Big(\sum_{k=1}^{K}\bm{W}_k^\star+\bm{W}_\mathrm{S}^\star\Big)-P\Big)=&0 \label{KKT_power}\\[-1mm]
	\mathrm{tr}(\bm{Z}_k^\star \bm{W}_k^\star)=&0,\quad \forall k \label{KKT2}\\[-1.5mm]
	\mathrm{tr}(\bm{Z}_\mathrm{S}^\star\bm{W}_\mathrm{S}^\star)=&0 \label{KKT3}\\[-1.5mm]
	\frac{\partial\mathcal{L}^\star}{\partial t^\star}=1-z_1^\star=&0 \label{derivative_1}\\[-1.5mm]
	\frac{\partial\mathcal{L}^\star}{\partial\bm{W}_k^\star}=&\bm{0},\quad\forall k \label{derivative_2}\\[-1mm]
	\frac{\partial\mathcal{L}^\star}{\partial\bm{W}_\mathrm{S}^\star}=&\bm{0}, \label{derivative_3}
\vspace{-2mm}
\end{align}
where (\ref{KKT1})-(\ref{KKT3}) are complementary slackness conditions and (\ref{derivative_1})-(\ref{derivative_3}) are first-order optimality conditions.

Next, we exploit the above KKT conditions. From (\ref{derivative_1}), we have $z^\star_1=1$, and consequently $\bm{Z}^\star_B \neq \bm{0}$. From (\ref{KKT1}), we have $|\bm{Z}_B^\star|=z^\star_1z^\star_3-|z^\star_2|^2=0$. Thus, we can obtain that $\bm{Z}_B^\star=[1,z_2^\star;z_2^{\star^*},|z_2^\star|^2]$, and $\bm{D}^\star$ can be expressed as $\bm{D}^\star=\bm{A}_1+z^\star_2\bm{A}^H_2+z_2^{\star^*}\bm{A}_2+|z^\star_2|^2\bm{A}_3=\int_{-\pi}^\pi \bar{\bm{D}}(\theta) p_\Theta(\theta)d\theta$, where
\begin{align}\label{matrixD}
	\vspace{-1mm}
	\!\!\bar{\bm{D}}(\theta)&=[\bm{a}(\phi,\theta),\dot{\bm{a}}(\phi,\theta)]\times\nonumber\\
	&\left[ \begin{array}{ll}
		\|\dot{\bm{b}}(\phi,\theta)\|^2+|z^\star_2|^2N_r & z^\star_2N_r\\
		z^{\star^*}_2 N_r & N_r\end{array}\right]
	\left[ \begin{array}{l}
		\bm{a}^H(\phi,\theta)\\
		\dot{\bm{a}}^H(\phi,\theta)\end{array}\right].
	\vspace{-1mm}
\end{align}
Since $\bar{\bm{D}}(\theta)$ can be shown to be a positive semi-definite matrix, we have $\bm{D}^\star\succeq\bm{0}$. Based on the above as well as Proposition \ref{prop_Communication_Beam}, we have the following proposition.
\begin{proposition}\label{prop_rank_sum}
	Given any optimal solution to (P2-R), we can always construct another optimal solution $(t^\star,\{\hat{\bm{W}}_k^\star\}_{k=1}^{K},\hat{\bm{W}}^\star_\mathrm{S})$ with $\hat{\bm{W}}_k^\star=\beta_kP\frac{\bm{W}^\star_k\bm{h}_k\bm{h}_k^H\bm{W}^\star_k}{\|\bm{W}^\star_k\bm{h}_k\|^2},\ \forall{k}$ and $\mathrm{rank}(\hat{\bm{W}}^\star_\mathrm{S})\leq1$, where $\beta_k\in(0,1]$ denotes the power allocation coefficient for user $k$ determined by the given optimal solution and satisfies $\sum_{k=1}^{K}\beta_k\leq1$.
\end{proposition}
\begin{IEEEproof}[Proof Sketch]
	We devise a tailored rank-reduction method where a new optimal solution with lower-ranked $\bm{W}_{\mathrm{S}}^\star$ can be constructed which achieves the same $\mathrm{PCRB}_\theta^{\mathrm{P}}$ (i.e., $[\bm{F}^{-1}]_{1,1}$) and communication rates, until $\mathrm{rank}(\hat{\bm{W}}^\star_\mathrm{S})\leq1$. Please refer to Appendix B for detailed proof.
\end{IEEEproof}

Therefore, \emph{at most one} dedicated sensing beam suffices to achieve the optimal sensing-communication trade-off in our considered multi-user ISAC system with heterogeneous unknown parameters, and a low-rank solution can be found via the rank-reduction method in the proof of Proposition \ref{prop_rank_sum}.

\begin{remark}\emph{(Difference between numbers of sensing beams needed with heterogeneous vs. homogeneous unknown parameters)}.
	It is worth noting that \cite{YaoJiayi2,Yu_Bound2} also derived bounds on the number of sensing beams needed for multi-user ISAC, under the assumption that $\theta$ and $\alpha$ are \emph{both random} parameters with \emph{known distributions}. With $\alpha$ treated as a nuisance parameter and sensing interference cancellation capability ignored, the number of sensing beams needed was shown to be $1$ in \cite{YaoJiayi2}, where the standard rank-reduction method in \cite{Palomar} was adopted. With $\alpha$ treated as a \emph{desired} parameter, the number of sensing beams needed was shown to be $2$ in \cite{Yu_Bound2}, where the rank-reduction method enforces all the elements in $\bm{F}^{-1}$ to be kept the same in each step to take into account the PFIM with respect to $\alpha$. In contrast, our tailored rank-reduction method only needs to ensure $[\bm{F}^{-1}]_{1,1}$ is kept the same, which may explain the tighter bound in this paper.
\end{remark}

\begin{figure}[t]
	\centering
	\includegraphics[width=3.5in]{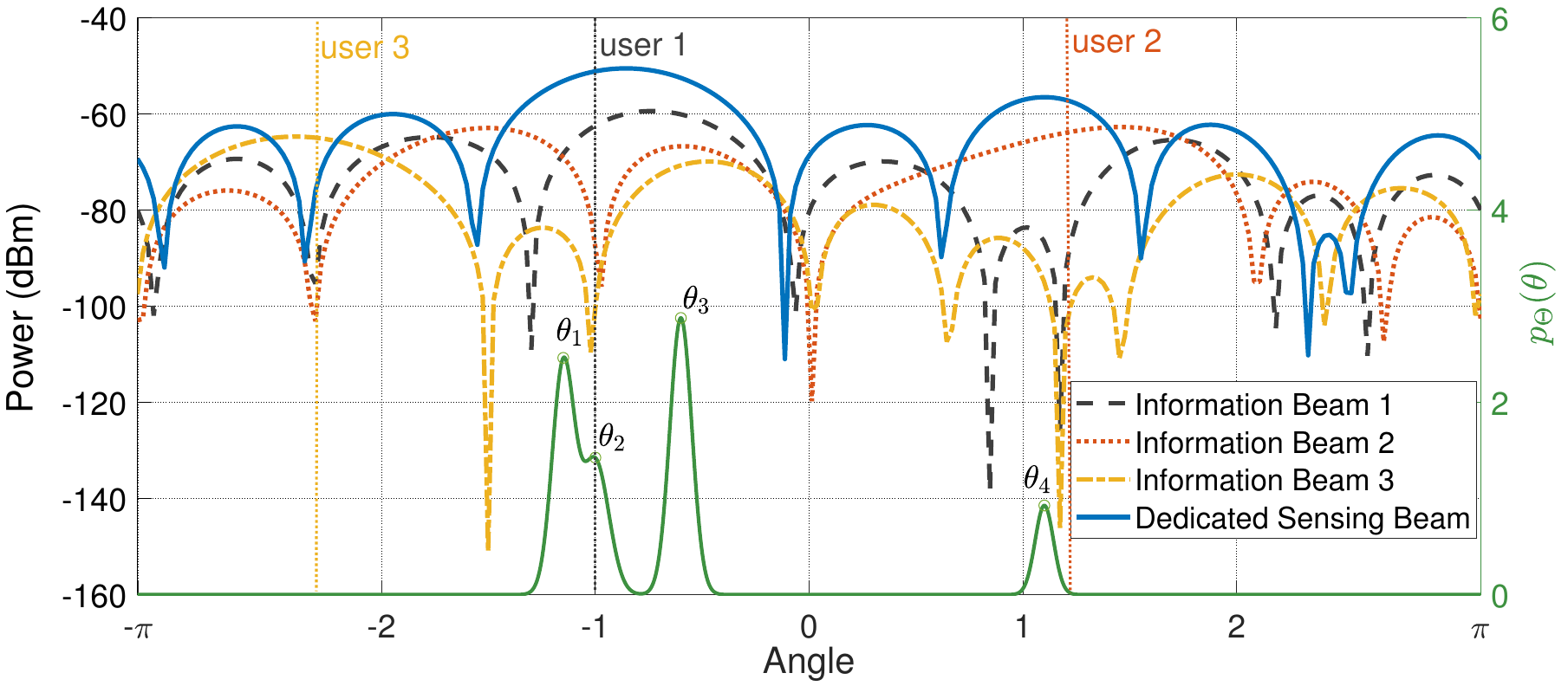}
	\vspace{-8mm}
	\caption{Illustration of radiation power pattern and target angle PDF $p_{\Theta}(\theta)$.}\label{Fig_Beampattern}
	\vspace{-1mm}
\end{figure}

\vspace*{-2mm}
\section{Numerical Results}\label{Sec_num}
\vspace{-1mm}
In this section, we provide numerical results to evaluate the performance of our proposed optimal beamforming design for multi-user ISAC with heterogeneous unknown parameters. We set $K\!=\!3$, $N_{\mathrm{T}}\!=\!3\times3$, $N_{\mathrm{R}}\!=\!3\times4$, $\sigma_\mathrm{S}^2\!=\!-90$ dBm, $r\!=\!50$ m, $\sigma_k^2\!=\!-90$ dBm, $\forall k$, and $\frac{P|\alpha|^2L}{\sigma_\mathrm{S}^2}\!=\!-5$ dB. We assume the BS height is $10$ m and the UPA has half-wavelength antenna spacing. We consider an LoS channel from the BS to each communication user. The heights, azimuth angles, and ranges of the users are set as $h_{\mathrm{U}_1}\!=\!h_{\mathrm{U}_2}\!=\!h_{\mathrm{U}_3}\!=\!1$ m; $\theta_{\mathrm{U}_1}\!=\!-1$, $\theta_{\mathrm{U}_2}\!=\!1.2$, $\theta_{\mathrm{U}_3}\!=\!-2.3$; and $r_{\mathrm{U}_1}\!=\!750$ m, $r_{\mathrm{U}_2}\!=\!650$ m, $r_{\mathrm{U}_3}\!=\!600$ m; respectively.
We adopt a von-Mises mixture PDF model given by $p_\Theta(\theta)\!=\!\sum_{m\!=\!1}^{M}p_m \frac{e^{\kappa_m\cos(\theta-\theta_m)}}{2\pi I_0(\kappa_m)}$ with $I_n(\kappa_m)\!=\!\frac{1}{2\pi}\int_{-\pi}^\pi e^{\kappa_m\cos t}\cos(nt)dt$ being the modified Bessel function; $M\!=\!4$; $\theta_1\!=\!-1.15$, $\theta_2\!=\!-1$, $\theta_3\!=\!-0.6$, $\theta_4\!=\!1.1$; $\kappa_1 \!=\!370$, $\kappa_2\!=\!250$, $\kappa_3\!=\!380$, $\kappa_4\!=\!540$; and $p_1\!=\!0.31$, $p_2\!=\!0.22$, $p_3\!=\!0.37$, $p_4\!=\!0.1$. $p_\Theta(\theta)$ is illustrated in Fig. \ref{Fig_Beampattern}.

First, we show the radiation power pattern at $r$ over different angles with the proposed solution in Fig. \ref{Fig_Beampattern} under $\bar{R}_k=4.5$ bps/Hz, $\forall k$. It is observed that the obtained radiation power patterns of information beams are concentrated over the corresponding user angle and target angles with high probability densities for assisting sensing. The radiation power pattern of the dedicated sensing beam is concentrated over target angles with high probability densities to enhance the strength of the echo signal, regardless of overlap with the user angles (e.g., user 1). This validates the effectiveness of our proposed design under sensing interference cancellation.

\begin{figure}[t]
	\centering
	\includegraphics[width=3.5in]{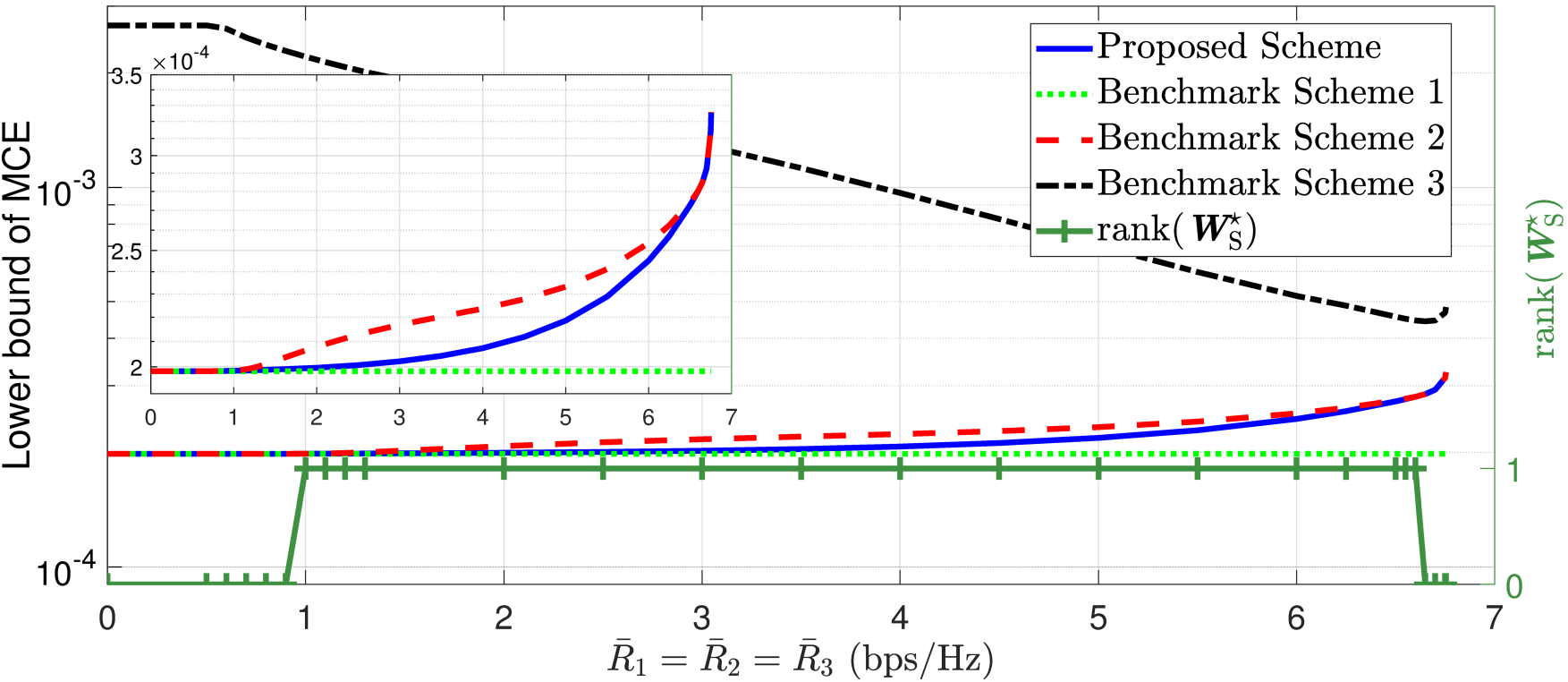}
	\vspace{-7mm}
	\caption{Lower bound of MCE versus communication rate target.}\label{Fig_PCRB_benchmark}
	\vspace{-1mm}
\end{figure}

Fig. \ref{Fig_PCRB_benchmark} shows the lower bound of MCE (periodic PCRB or expected periodic CRB) versus the communication rate target for the proposed design and the following benchmark schemes, where $\mathrm{rank}(\bm{W}_\mathrm{S}^\star)$ in our proposed design, i.e., the number of dedicated sensing beams, is also shown.
\vspace*{-1mm}
\begin{itemize}
	\item {\bf{Benchmark Scheme 1: Sensing-oriented beamforming design.}} The sensing beamforming matrix $\bm{S}$ is designed to minimize the periodic PCRB under the power constraint.
	\item {\bf{Benchmark Scheme 2: Dual-functional beamforming design.}} The dual-functional beamforming vectors, i.e., $\{\bm{w}_k\}_{k=1}^K$, are designed by solving (P1) with $\bm{S}=\bm{0}$.
	\item {\bf{Benchmark Scheme 3: Most-probable angle based beamforming design.}} Both $\{\bm{w}_k\}_{k=1}^K$ and $\bm{S}$ are designed to minimize the periodic CRB corresponding to the angle with the highest probability under the constraints in (P1).
\end{itemize}
\vspace*{-1mm}
It is observed that Benchmark Scheme 3 has significantly inferior performance to other schemes especially in the low-rate regime, due to the lack of full exploitation of the prior distribution information. Then, it is observed that our proposed solution requires at most one dedicated sensing beam, which is consistent with our analytical results in Section \ref{Sec_sol}. Moreover, our proposed scheme significantly outperforms Benchmark Scheme 2 in a wide range of rate regimes, which is also evidenced by the one sensing beam needed in this regime. This thus validates the need of dedicated sensing beam for optimizing the communication-sensing trade-off. Finally, our proposed scheme achieves close sensing performance to the sensing-only Benchmark Scheme 1, while the latter can only support extremely low communication rate.

\vspace*{-2mm}
\section{Conclusions}
\vspace*{-1mm}
This paper studied a multi-antenna ISAC system with heterogeneous unknown parameters and sensing interference cancellation capability at the communication users. With the aim of minimizing the periodic PCRB for sensing the desired unknown angle parameter under individual communication user rate constraints, we formulated the transmit beamforming optimization problem considering a general structure with both communication beams and dedicated sensing beams, for which the optimal solution was derived. Moreover, the number of dedicated sensing beams needed was analytically proved to be no larger than one. Numerical results verified our analysis and effectiveness of our proposed design.

\newpage
\newpage
\bibliographystyle{IEEEtran}
\bibliography{reference}

\newpage
\newpage
\appendices
\section{Proof of Proposition \ref{prop_Communication_Beam}}\label{proof_Communication_Beam}
Given any  optimal solution $(t^\star,\{\bm{W}^\star_k\}_{k=1}^{K},\bm{W}^\star_\mathrm{S})$, we can construct a new solution $(t^\star,\{\bar{\bm{W}}^\star_k\}_{k=1}^{K},\bar{\bm{W}}^\star_\mathrm{S})$, where $\bar{\bm{W}}^\star_k$ and $\bar{\bm{W}}^\star_\mathrm{S}$ are shown in (\ref{construct_Wu}) and (\ref{construct_WS}).

From $\sum_{k=1}^{K}\bar{\bm{W}}^\star_k+\bar{\bm{W}}^\star_\mathrm{S}=\sum_{k=1}^{K}\bm{W}^\star_k+\bm{W}^\star_\mathrm{S}$, constraints (\ref{P2R_C1}) and (\ref{P2R_C4}) hold for the new constructed solution.

For any vector $\bm{\omega}$, we have
\begin{align} \label{Cauchy}
	&\bm{\omega}^H(\bm{W}^\star_k-\bar{\bm{W}}^\star_k)\bm{\omega}\nonumber\\
	&=\frac{\bm{\omega}^H\bm{W}^\star_k\bm{\omega}\bm{h}_k^H\bm{W}^\star_k\bm{h}_k-|\bm{\omega}^H\bm{W}^\star_k\bm{h}_k|^2}{\bm{h}_k^H\bm{W}^\star_k\bm{h}_k}\geq0,
\end{align}
where the inequality holds due to Cauchy-Schwarz inequality. Based on (\ref{Cauchy}), we have $\bm{W}^\star_k-\bar{\bm{W}}^\star_k\succeq\bm{0},\ \forall k$. From $\bm{h}_k^H\bm{W}^\star_k\bm{h}_k=\bm{h}_k^H\bar{\bm{W}}^\star_k\bm{h}_k$ and $\bar{\bm{W}}^\star_\mathrm{S}-\bm{W}^\star_\mathrm{S}=\sum_{k=1}^{K}\bm{W}^\star_k-\sum_{k=1}^{K}\bar{\bm{W}}^\star_k\succeq\bm{0}$, we have
\begin{align}\label{SINR_TypeII}
	&\bm{h}_k^H(\bar{\bm{W}}^\star_k-\gamma_k\sum_{i\neq k}\bar{\bm{W}}^\star_i)\bm{h}_k\nonumber\\ &=\bm{h}_k^H((1+\gamma_k)\bm{W}^\star_k-\gamma_k\sum_{i=1}^K\bm{W}^\star_i+\gamma_k(\bar{\bm{W}}^\star_\mathrm{S}-\bm{W}^\star_\mathrm{S}))\bm{h}_k\nonumber\\
	&\geq\bm{h}_k^H((1\!+\!\gamma_k)\bm{W}^\star_k\!-\!\gamma_k\sum_{i=1}^K\bm{W}^\star_i)\bm{h}_k\!\geq\!\gamma_k\sigma_k^2,\quad\forall k.
\end{align}
Thus, (\ref{P2R_C3}) holds for $(\{\bar{\bm{W}}^\star_k\}_{k=1}^{K},\bar{\bm{W}}^\star_\mathrm{S})$.
Then, constraint (\ref{P2R_C5}) is satisfied by $\bar{\bm{W}}^\star_k\succeq\bm{0},\ \forall k$, and  constraint (\ref{P2R_C6}) is satisfied by $\bar{\bm{W}}^\star_\mathrm{S}\succeq\bm{0}$ due to that $\bm{W}^\star_k-\bar{\bm{W}}^\star_k\succeq\bm{0},\ \forall k$ holds.

The optimal objective value of (P2-R) achieved by $(t^\star,\{\bar{\bm{W}}^\star_k\}_{k=1}^{K},\bar{\bm{W}}^\star_{\mathrm{S}})$ is the same as $(t^\star,\{\bm{W}^\star_k\}_{k=1}^{K},\bm{W}^\star_\mathrm{S})$, and all the constraints in (P2-R) are satisfied by $(t^\star,\{\bar{\bm{W}}^\star_k\}_{k=1}^{K},\bar{\bm{W}}^\star_{\mathrm{S}})$. Therefore, $(t^\star,\{\bar{\bm{W}}^\star_k\}_{k=1}^{K},\bar{\bm{W}}^\star_{\mathrm{S}})$ is an optimal solution to (P2-R). Proposition \ref{prop_Communication_Beam} is thus proved.

\section{Proof of Proposition \ref{prop_rank_sum}}\label{proof_rank_sum}
Given any optimal transmit covariance matrix $\bm{R}_X^\star=\sum_{k=1}^{K}\bm{W}^\star_k+\bm{W}^\star_\mathrm{S}$, we have the following lemma.
\begin{lemma}\label{lemma1}
	$\bm{R}_X^\star$ satisfies that $t^\star=\mathrm{tr}(\bm{D}^\star\bm{R}_X^\star)$ and $\mathrm{tr}((\bm{A}_2+z_2^\star\bm{A}_3)\bm{R}_X^\star)=0$.
\end{lemma}
\begin{IEEEproof}
	Based on  condition (\ref{KKT1}), we have $t^\star=\mathrm{tr}(\bm{D}^\star\bm{R}_X^\star)$. Also, from $\bm{Z}^\star_B\neq\bm{0}$ and  condition (\ref{KKT1}), we can obtain that $|\bm{B}\left(t^\star,\{\bm{W}_k^\star\}_{k=1}^{K},\bm{W}^\star_\mathrm{S}\right)|=0$, which implies that $t^\star=\mathrm{tr}(\bm{A}_1\bm{R}_X^\star)-\frac{|\mathrm{tr}(\bm{A}_2\bm{R}_X^\star)|^2}{\mathrm{tr}(\bm{A}_3\bm{R}_X^\star)}$. Thus, we have
	\begin{align}\label{DW}
		t^\star&=\mathrm{tr}((\bm{A}_1+z^\star_2\bm{A}^H_2+z_2^{\star^*}\bm{A}_2+|z^\star_2|^2\bm{A}_3)\bm{R}_X^\star)\nonumber\\
		&=\mathrm{tr}(\bm{A}_1\bm{R}_X^\star)-\frac{|\mathrm{tr}(\bm{A}_2\bm{R}_X^\star)|^2}{\mathrm{tr}(\bm{A}_3\bm{R}_X^\star)}.
	\end{align}
	Based on (\ref{DW}), we can obtain that
	\begin{align}
		&(z_2^\star\mathrm{tr}(\bm{A}_3\bm{R}_X^\star)+\mathrm{tr}(\bm{A}_2\bm{R}_X^\star))(z_2^{\star^*}\mathrm{tr}(\bm{A}_3\bm{R}_X^\star)+\mathrm{tr}(\bm{A}_2^H\bm{R}_X^\star))\nonumber\\
		&=|\mathrm{tr}((\bm{A}_2+z_2^\star\bm{A}_3)\bm{R}_X^\star)|^2=0,
	\end{align}
	which yields that $\mathrm{tr}((\bm{A}_2+z_2^\star\bm{A}_3)\bm{R}_X^\star)=0$.
\end{IEEEproof}

For any transmit covariance matrix $\bm{R}_X\succeq\bm{0}$, we have the following Lemma.
\begin{lemma}\label{lemma2}
	If $\mathrm{tr}((\bm{A}_2+z_2^\star\bm{A}_3)\bm{R}_X)=0$, we have $\left[\begin{array}{ll}\mathrm{tr}(\bm{A}_1\bm{R}_X)-\mathrm{tr}(\bm{D}^\star\bm{R}_X)&\mathrm{tr}(\bm{A}_2\bm{R}_X)\\
		\mathrm{tr}(\bm{A}_2^H\bm{R}_X)&\mathrm{tr}(\bm{A}_3\bm{R}_X)\end{array}\right]\succeq\bm{0}$.
\end{lemma}
\begin{IEEEproof}
	From $\mathrm{tr}((\bm{A}_2+z_2^\star\bm{A}_3)\bm{R}_X)=0$, we have
	\begin{align}\label{lemma1_equ}
		&|\mathrm{tr}((\bm{A}_2+z_2^\star\bm{A}_3)\bm{R}_X)|^2\nonumber\\
		&=(z_2^\star\mathrm{tr}(\bm{A}_3\bm{R}_X)\!+\!\mathrm{tr}(\bm{A}_2\bm{R}_X))(z_2^{\star^*}\mathrm{tr}(\bm{A}_3\bm{R}_X)\!+\!\mathrm{tr}(\bm{A}_2^H\bm{R}_X))\nonumber\\
		&=0.
	\end{align}
	In this work, we consider $\mathrm{tr}(\bm{A}_3\bm{R}_X)>0$, and it follows (\ref{lemma1_equ}) that
	\begin{align}
		\mathrm{tr}(\bm{A}_1\bm{R}_X)-\mathrm{tr}(\bm{D}^\star\bm{R}_X)\!-\!\frac{|\mathrm{tr}(\bm{A}_2\bm{R}_X)|^2}{\mathrm{tr}(\bm{A}_3\bm{R}_X)}=0.
	\end{align}
	Based on the Schur complement condition \cite{schur}, $\mathrm{tr}(\bm{A}_1{\bm{R}}_X)-\mathrm{tr}(\bm{D}^\star\bm{R}_X)-\frac{|\mathrm{tr}(\bm{A}_2{\bm{R}}_X)|^2}{\mathrm{tr}(\bm{A}_3{\bm{R}}_X)}=0$ with $\mathrm{tr}(\bm{A}_3{\bm{R}}_X)>0$ is equivalent to $\left[\begin{array}{ll}\mathrm{tr}(\bm{A}_1\bm{R}_X)-\mathrm{tr}(\bm{D}^\star\bm{R}_X)&\mathrm{tr}(\bm{A}_2\bm{R}_X)\\
		\mathrm{tr}(\bm{A}_2^H\bm{R}_X)&\mathrm{tr}(\bm{A}_3\bm{R}_X)\end{array}\right]\succeq\bm{0}$.
\end{IEEEproof}

Given optimal solution $(t^\star,\{\bar{\bm{W}}_k^\star\}_{k=1}^{K},\bar{\bm{W}}^\star_\mathrm{S})$ obtained in Proposition \ref{prop_Communication_Beam}, let $M_k=\mathrm{rank}(\bar{\bm{W}}_k^\star)=1,\ \forall k$ and $M_\mathrm{S}=\mathrm{rank}(\bar{\bm{W}}_\mathrm{S}^\star)$. We express $\bar{\bm{W}}_k^\star=\bm{f}_k\bm{f}_k^H$ with $\bm{f}_k\in \mathbb{C}^{N_\mathrm{T}\times 1}$ for each $k$ and $\bar{\bm{W}}_{\mathrm{S}}^\star=\bm{J}_\mathrm{S}\bm{J}_\mathrm{S}^H$ with $\bm{J}_\mathrm{S}\in \mathbb{C}^{N_\mathrm{T}\times M_\mathrm{S}}$. Then, we introduce $K$ real values $\Delta_k,\ k=1,...,K$ and one Hermitian matrix $\bm{\Delta}_\mathrm{S}=\bm{\Delta}_{\mathrm{S}_\mathrm{R}}+j\bm{\Delta}_{\mathrm{S}_\mathrm{I}}\in \mathbb{C}^{M_\mathrm{S}\times M_\mathrm{S}}$. Since $\bm{\Delta}_\mathrm{S}$ is a Hermitian matrix, there are $\frac{M_\mathrm{S}^2+M_\mathrm{S}}{2}$ and $\frac{M_\mathrm{S}^2-M_\mathrm{S}}{2}$ unknown real values in $\bm{\Delta}_{\mathrm{S}_\mathrm{R}}$ and $\bm{\Delta}_{\mathrm{S}_\mathrm{I}}$, respectively. Therefore, there are $M_\mathrm{S}^2$ unknown real values in $\bm{\Delta}_\mathrm{S}$.

Then, we consider the following system of linear equations:
\begin{align}
	&\mathfrak{Re}\Big\{\sum_{k=1}^{K}\bm{f}_k^H(\bm{A}_2+z_2^\star\bm{A}_3)\bm{f}_k{\Delta}_k\Big\}\nonumber\\
	&\quad\quad+\mathfrak{Re}\Big\{\mathrm{tr}(\bm{J}_\mathrm{S}^H(\bm{A}_2+z_2^\star\bm{A}_3)\bm{J}_\mathrm{S}\bm{\Delta}_\mathrm{S})\Big\}=0,\label{linear1}\\
	&\mathfrak{Im}\Big\{\sum_{k=1}^{K}\bm{f}_k^H(\bm{A}_2+z_2^\star\bm{A}_3)\bm{f}_k{\Delta}_k\Big\}\nonumber\\
	&\quad\quad+\mathfrak{Im}\Big\{\mathrm{tr}(\bm{J}_\mathrm{S}^H(\bm{A}_2+z_2^\star\bm{A}_3)\bm{J}_\mathrm{S}\bm{\Delta}_\mathrm{S})\Big\}=0,\label{linear2}\\
	&|\bm{f}_k^H\bm{h}_k|^2{\Delta}_k-\gamma_k\sum_{i\neq{k}}|\bm{f}_i^H\bm{h}_k|^2{\Delta}_i=0,\quad \forall k,\label{linear4}\\
	&\sum_{k=1}^{K}\bm{f}_k^H\bm{f}_k{\Delta}_k+\mathrm{tr}(\bm{J}_\mathrm{S}^H\bm{J}_\mathrm{S}\bm{\Delta}_\mathrm{S})=0.\label{linear5}
\end{align}
In equations (\ref{linear1})-(\ref{linear5}), there are $K+3$ equations and $\sum_{k=1}^{K}M_k^2+M_\mathrm{S}^2=K+M_\mathrm{S}^2$ unknown real values.

If $\sum_{k=1}^{K}M_k^2+M_\mathrm{S}^2>K+3$, i.e. $M_\mathrm{S}^2>3$, then there is a non-zero solution of the system of linear (\ref{linear1})-(\ref{linear5}). Let $\xi_1,...,\xi_{M_\mathrm{S}}$ denote the eigenvalues of matrix $\bm{\Delta}_\mathrm{S}$. Define $\hat{\xi}$ as a value in $\{\Delta_1,...\Delta_{K},\xi_1,...,\xi_{M_\mathrm{S}}\}$ with maximum absolute value, where  $|\hat{\xi}|=\max\{|\Delta_1|,...|\Delta_{K}|,|\xi_1|,...,|\xi_{M_\mathrm{S}}|\}$.
Based on $\hat{\xi}$, we can construct matrices $\hat{\bm{W}}_k$'s and $\hat{\bm{W}}_\mathrm{S}$ as
\begin{align}
	&\hat{\bm{W}}_k=(1\!-\!\Delta_k/\hat{\xi})\bm{f}_k\bm{f}_k^H=(1\!-\!\Delta_k/\hat{\xi})\bar{\bm{W}}_k\succeq\bm{0},\quad \forall k,\label{Wu_new}\\
	&\hat{\bm{W}}_\mathrm{S}=\bm{J}_\mathrm{S}(\bm{I}_{M_\mathrm{S}}-\bm{\Delta}_\mathrm{S}/\hat{\xi})\bm{J}_\mathrm{S}^H\succeq\bm{0}.\label{WS_new}
\end{align}
Then, we prove that  $(t^\star,\{\hat{\bm{W}}_k\}_{k=1}^{K},\hat{\bm{W}}_\mathrm{S})$ is also an optimal solution to (P2-R).

From (\ref{linear4}), we have
\begin{align}
	&\bm{h}_k^H(\hat{\bm{W}}_k-\gamma_k\sum_{i\neq k}\hat{\bm{W}}_i)\bm{h}_k\nonumber\\
	&=\bm{h}_k^H(\bar{\bm{W}}_k^\star-\gamma_k\sum_{i\neq k}\bar{\bm{W}}_i^\star)\bm{h}_k-\Big(|\bm{f}_k^H\bm{h}_k|^2{\Delta}_k\nonumber\\
	&\quad\quad-\gamma_k\sum_{i\neq k}|\bm{f}_i^H\bm{h}_k|^2{\Delta}_i\Big)/\hat{\xi}\geq\gamma_k\sigma^2_k,\quad\forall k.
\end{align}
Thus, we can know that the rate constraints of communication users in (\ref{P2R_C3}) are satisfied by the constructed solution.

From (\ref{linear5}), we have
\begin{align}\label{equ_power}
	\mathrm{tr}\Big(\sum_{k=1}^{K}\hat{\bm{W}}_k+\hat{\bm{W}}_\mathrm{S}\Big)&=\mathrm{tr}\Big(\sum_{k=1}^{K}\bm{f}_k\bm{f}_k^H+\bm{J}_\mathrm{S}\bm{J}_\mathrm{S}^H\Big)\nonumber\\
	&-\Big(\sum_{k=1}^{K}\bm{f}_k^H\bm{f}_k{\Delta}_k+\mathrm{tr}(\bm{J}_\mathrm{S}^H\bm{J}_\mathrm{S}\bm{\Delta}_\mathrm{S})\Big)/\hat{\xi}\nonumber\\
	&=\mathrm{tr}\Big(\sum_{k=1}^{K}\bar{\bm{W}}_k^\star+\bar{\bm{W}}_\mathrm{S}^\star\Big),
\end{align}
based on which the power constraint (\ref{P2R_C4}) is satisfied.

Moreover, since the communication rate constraints are satisfied by $\{\hat{\bm{W}}_k\}_{k=1}^{K}$, we can know that $\hat{\bm{W}}_k\neq\bm{0},\ \forall k$ holds. Thus, we have $|\hat{\xi}|=\max\{|\xi_1|,...,|\xi_{M_\mathrm{S}}|\}$, which implies that $|\bm{I}_{M_\mathrm{S}}-\bm{\Delta}_\mathrm{S}/\hat{\xi}|=0$ and $\mathrm{rank}(\hat{\bm{W}}_\mathrm{S})\leq\mathrm{rank}(\bar{\bm{W}}_\mathrm{S}^\star)-1$.

Then, define $\hat{\bm{R}}_X=\sum_{k=1}^{K}\hat{\bm{W}}_k+\hat{\bm{W}}_\mathrm{S}$ and $\bar{\bm{R}}_X=\sum_{k=1}^{K}\bar{\bm{W}}_k^\star+\bar{\bm{W}}^\star_\mathrm{S}$. From (\ref{linear1}), (\ref{linear2}), and Lemma \ref{lemma1}, we have
\begin{align}\label{C1_linear1}
	\mathrm{tr}((\bm{A}_2+z_2^\star\bm{A}_3)\hat{\bm{R}}_X)=\mathrm{tr}((\bm{A}_2+z_2^\star\bm{A}_3)\bar{\bm{R}}_X)=0.
\end{align}

Based on KKT conditions (\ref{derivative_2}) and (\ref{derivative_3}), we have
\begin{align}
	&\bm{Z}_k^{\star}=\eta^\star\bm{I}_{N_\mathrm{T}}-\bm{D}^\star+\sum_{i\neq{k}}\mu_i^\star\gamma_i\bm{h}_i\bm{h}_i^H-\mu_k^\star\bm{h}_k\bm{h}_k^H,\;\forall{k},\label{Zu}\\
	&\bm{Z}_\mathrm{S}^{\star}=\eta^\star\bm{I}_{N_\mathrm{T}}-\bm{D}^\star.\label{ZS}
\end{align}
Since the optimal solution satisfies the conditions (\ref{KKT2}) and (\ref{KKT3}), we have
\begin{align}
	&\mathrm{tr}(\bm{Z}^\star_k\bar{\bm{W}}_k^\star)=\eta^\star\mathrm{tr}(\bar{\bm{W}}_k^\star)-\mathrm{tr}(\bm{D}^\star\bar{\bm{W}}_k^\star)\nonumber\\
	&\quad+\mathrm{tr}(\bar{\bm{W}}^\star_k(\sum_{i\neq{k}}\mu_i^\star\gamma_i\bm{h}_i\bm{h}_i^H-\mu_k^\star\bm{h}_k\bm{h}_k^H))=0,\quad\forall{k},\label{ZuWu}\\
	&\mathrm{tr}(\bm{Z}^\star_\mathrm{S}\bar{\bm{W}}^\star_\mathrm{S})=\eta^\star\mathrm{tr}(\bar{\bm{W}}^\star_\mathrm{S})-\mathrm{tr}(\bm{D}^\star\bar{\bm{W}}^\star_\mathrm{S})=0.\label{ZSWS}
\end{align}
The summation of (\ref{ZuWu}) and (\ref{ZSWS}) is given by
\begin{align}\label{sum1}
	&\sum_{k=1}^{K}\mathrm{tr}(\bm{Z}^\star_k\bar{\bm{W}}^\star_k)+\mathrm{tr}(\bm{Z}^\star_\mathrm{S}\bar{\bm{W}}^\star_\mathrm{S})=\eta^\star\mathrm{tr}(\bar{\bm{R}}_X)-\mathrm{tr}(\bm{D}^\star\bar{\bm{R}}_X)\nonumber\\
	&\quad-\sum_{k=1}^{K}\mu_k^\star\bm{h}_k^H(\bar{\bm{W}}_k^\star-\gamma_k\sum_{i\neq{k}}\bar{\bm{W}}^\star_i)\bm{h}_k=0.
\end{align}
On the other hand, we have $\mathrm{tr}(\bm{Z}^\star_k\hat{\bm{W}}_k)=(1-\Delta_k/\hat{\xi})\mathrm{tr}(\bm{Z}^\star_k\bar{\bm{W}}^\star_k)=0$.
From $\mathrm{tr}(\bm{Z}^\star_\mathrm{S}\bar{\bm{W}}^\star_\mathrm{S})=\mathrm{tr}(\bm{J}_\mathrm{S}^H\bm{Z}^\star_\mathrm{S}\bm{J}_\mathrm{S})=0$, we have $\bm{J}_\mathrm{S}^H\bm{Z}^\star_\mathrm{S}\bm{J}_\mathrm{S}=\bm{0}$. Thus, we can obtain that $\mathrm{tr}(\bm{Z}^\star_\mathrm{S}\hat{\bm{W}}_\mathrm{S})=\mathrm{tr}(\bm{J}_\mathrm{S}^H\bm{Z}^\star_\mathrm{S}\bm{J}_\mathrm{S}(\bm{I}_{M_\mathrm{S}}-\bm{\Delta}_\mathrm{S}/\hat{\xi}))=0$. Similar, based on $\mathrm{tr}(\bm{Z}^\star_k\hat{\bm{W}}_k)=0$ and $\mathrm{tr}(\bm{Z}^\star_\mathrm{S}\hat{\bm{W}}_\mathrm{S})=0$, we have
\begin{align}\label{sum2}
	&\sum_{k=1}^{K}\mathrm{tr}(\bm{Z}^\star_k\hat{\bm{W}}_k)+\mathrm{tr}(\bm{Z}^\star_\mathrm{S}\hat{\bm{W}}_\mathrm{S})=\eta^\star\mathrm{tr}(\hat{\bm{R}}_X)-\mathrm{tr}(\bm{D}^\star\hat{\bm{R}}_X)\nonumber\\
	&\quad-\sum_{k=1}^{K}\mu_k^\star\bm{h}_k^H(\hat{\bm{W}}_k-\gamma_k\sum_{i\neq{k}}\hat{\bm{W}}_i)\bm{h}_k=0.
\end{align}
From (\ref{sum1}), (\ref{sum2}), and Lemma 1,  we can obtain that
\begin{align}\label{C1_linear2}
	\mathrm{tr}(\bm{D}^\star\hat{\bm{R}}_X)=\mathrm{tr}(\bm{D}^\star\bar{\bm{R}}_X)=t^\star.
\end{align}
From (\ref{C1_linear1}), (\ref{C1_linear2}) and Lemma \ref{lemma2}, we can know that constraint (\ref{P2R_C1}) in (P2-R) is satisfied by $\bm{B}\left(t^\star,\{\hat{\bm{W}}_k\}_{k=1}^{K},\hat{\bm{W}}_\mathrm{S}\right)=\left[\begin{array}{ll}
	\mathrm{tr}(\bm{A}_1\hat{\bm{R}}_X)-t^\star&\mathrm{tr}(\bm{A}_2\hat{\bm{R}}_X)\\
	\mathrm{tr}(\bm{A}_2^H\hat{\bm{R}}_X)&\mathrm{tr}(\bm{A}_3\hat{\bm{R}}_X)
\end{array}\right]=\left[\begin{array}{ll}
	\mathrm{tr}(\bm{A}_1\hat{\bm{R}}_X)-\mathrm{tr}(\bm{D}^\star\hat{\bm{R}}_X)&\mathrm{tr}(\bm{A}_2\hat{\bm{R}}_X)\\
	\mathrm{tr}(\bm{A}_2^H\hat{\bm{R}}_X)&\mathrm{tr}(\bm{A}_3\hat{\bm{R}}_X)
\end{array}\right]\succeq \bm{0}$. Furthermore, the constraints (\ref{P2R_C5}) and (\ref{P2R_C6}) are also satisfied by $\hat{\bm{W}}_k\succeq\bm{0},\ \forall k$ and $\hat{\bm{W}}_{\mathrm{S}}\succeq\bm{0}$.

It is shown that the optimal objective value of (P2-R) achieved by $(t^\star,\{\hat{\bm{W}}_k\}_{k=1}^{K},\hat{\bm{W}}_{\mathrm{S}})$ is the same as the optimal solution $(t^\star,\{\bar{\bm{W}}_k^\star\}_{k=1}^{K},\bar{\bm{W}}^\star_{\mathrm{S}})$, and all the constraints in (P2-R) are satisfied by $(t^\star,\{\hat{\bm{W}}_k\}_{k=1}^{K},\hat{\bm{W}}_{\mathrm{S}})$. Therefore, $(t^\star,\{\hat{\bm{W}}_k\}_{k=1}^{K},\hat{\bm{W}}_{\mathrm{S}})$ is an optimal solution to (P2-R).

As we have proved that $\mathrm{rank}(\hat{\bm{W}}_\mathrm{S})\leq\mathrm{rank}(\bar{\bm{W}}_\mathrm{S}^\star)-1$, we can repeat the above rank-deduction procedure until we finally find an optimal solution $(t^\star,\{\hat{\bm{W}}_k^\star\}_{k=1}^{K},\hat{\bm{W}}^\star_{\mathrm{S}})$ with $\sum_{k=1}^{K}\mathrm{rank}^2(\hat{\bm{W}}^\star_k)+\mathrm{rank}^2(\hat{\bm{W}}^\star_{\mathrm{S}})=K+\mathrm{rank}^2(\hat{\bm{W}}^\star_{\mathrm{S}})\leq K+3$ \cite{Palomar}. Thus, we can obtain that $\mathrm{rank}^2(\hat{\bm{W}}^\star_{\mathrm{S}})\leq3$, i.e. $\mathrm{rank}(\hat{\bm{W}}^\star_{\mathrm{S}})\leq1$.

From equation (\ref{Wu_new}), we can know that $\hat{\bm{W}}^\star_k$ and $\bar{\bm{W}}_k^\star$ have the same eigenvector, which is $\bm{W}_k^\star\bm{h}_k/\|\bm{W}_k^\star\bm{h}_k\|$. Thus, the rank-one structure of $\hat{\bm{W}}^\star_k$ is $\hat{\bm{W}}^\star_k=\beta_kP\frac{\bm{W}^\star_k\bm{h}_k\bm{h}_k^H\bm{W}^\star_k}{\|\bm{W}^\star_k\bm{h}_k\|^2}$ with $\beta_k\in(0,1]$. From $\mathrm{tr}\Big(\sum_{k=1}^{K}\hat{\bm{W}}^\star_k\Big)\geq P$, we have $\sum_{k=1}^{K}\beta_k\leq1$. Proposition \ref{prop_rank_sum} is thus proved.
\end{document}